\documentclass[fleqn,11pt]{wlscirep}
\usepackage[utf8]{inputenc}
\usepackage[T1]{fontenc}
\usepackage{kotex}
\usepackage{setspace}
\title{GCIceNet: A Graph Convolutional Network for Accurate Classification of Water Phases}

\author[1]{QHwan Kim}
\author[1]{Joon-Hyuk Ko}
\author[1]{Sunghoon Kim}
\author[1*]{Wonho Jhe}
\affil[1]{Department of Physics and Astronomy, Institute of Applied Physics, Seoul National University, Gwanak-gu, Seoul 08826, Republic of Korea.}

\affil[*]{whjhe@snu.ac.kr}


\begin{abstract}
Understanding phases of water molecules based on local structure is essential for understanding their anomalous properties. However, due to complicated structural motifs formed via hydrogen bonds, conventional order parameters represent the water molecules incompletely. In this paper, we develop a GCIceNet, which automatically generates machine-based order parameters for classifying the phases of the water molecules via supervised and unsupervised learning. Multiple graph convolutional layers in the GCIceNet can learn topological informations of the complex hydrogen bond networks. It shows a substantial improvement of accuracy for predicting the phase of water molecules in the bulk system and the ice/vapor interface system. A relative importance analysis shows that the GCIceNet can capture the structural features of the given system hidden in the input data. Augmented with the vast amount of data provided by molecular dynamics simulations, the GCIceNet is expected to serve as a powerful tool for the fields of glassy liquids and hydration layers around biomolecules.
\end{abstract}
\begin{document}

\flushbottom
\maketitle
%
%

\section*{Introduction}

A hydrogen bond between water molecules promote the formation of various crystalline structures of ice, including at least 17 polymorphs, in nature\cite{Salzmann2019}. A structural motif of the hydrogen bond network determines inherent proton ordering structures with the long-range order\cite{Loubeyre1999, Nanda2013, Komatsu2016}, thermal expansion coefficient\cite{Rottger1994, Noya2007}, dielectric spectrum\cite{Plaga2019}, and self-diffusion of protons\cite{Noguchi2016} of the ice. These studies have been used in ice-related studies such as ice discovery on Mars\cite{Dundas2018}, ice precipitation in cloud\cite{Moberg2019}, and a formation of gas-filled clathrate hydrates\cite{Loveday2001, Schaack2019}. Therefore, the development of an order parameter that represents the molecular fingerprint of the ice polymorphs is a crucial topic in the study on water.

From the position and momentum of atoms obtained from molecular dynamics simulation, many studies have developed the order parameters focused on a certain property of molecular arrangement, such as the tetrahedral\cite{Errington2001}, translational\cite{Shiratani1998}, and symmetrical properties\cite{Steinhardt1983, Martelli2018} of the water molecules\cite{Tanaka2019}. These parameters have been used to define the structural properties of ice\cite{Steinhardt1983}, supercooled water\cite{Errington2001, Kumar2009},  hydration layers\cite{Duboue-Dijon2015}. However, due to the complicated mathematical expressions involved, most order parameters reflect only partial information about the arrangement of the water molecules. The selection of the order parameter that gives the best description of a given system is usually a time-consuming process\cite{Duboue-Dijon2015}, sometimes even requiring the development of a more refined parameter\cite{Russo2014}. Furthermore, in complex systems where several symmetries coexist, the single order parameter is insufficient to describe the system, and a combination of the parameters should be used instead\cite{Lupi2017}.

Recently, data-driven approaches based on deep learning have been proposed to describe the properties of the complex system more accurately than the traditional approaches. In this approach, the order parameter is represented by the nonlinear mapping between input data and desired output labels with minimum loss\cite{Carrasquilla2017}. An optimization strategy from a backpropagation algorithm with multiple neural network layers can find the best mappings without the human-made complex mathematics. The predictability of the deep learning outperforms hand-made order parameters in the field of image classification\cite {Krizhevsky2012, Simonyan2014, Lecun2015}, natural language processing\cite{Bahdanau2014, Sutskever2014}, speech recognition\cite{Sak2015, Amodei2016}, and condensed matter physics\cite{Carrasquilla2017}.
 
In this letter, we use a deep learning-based approach to develop the data-driven order parameters of the water molecules. We introduce a GCIceNet with the graph convolutional networks. As such, graph convolutional networks has been used to study on citation networks \cite{Kipf2016}, structure-property relationships of organic molecules\cite{Schutt2017, Wu2018, Lim2019}, inorganic crystals\cite{Xie2018}, and design of organic molecules\cite{Cao2018, Bresson2019}.  We use the graph convolutional networks because the configurations of water molecules can be well accommodated with the graph representation, with nodes being the individual water molecules, and edges being the hydrogen bonds. Also, each node carries a feature vector that encodes further information containing the structural properties of the water molecule. With the graph data, the GCIceNet can learn the mapping from the given molecular graph to the water phase, which then can be used to classify water molecules by their phases. The accuracy of classifying the different ice phases with the GCIceNet outperforms other traditional methods. 
 
 To quantify the performance of the GCIceNet, we carry out the molecular dynamics simulation of two kinds of systems: i) a single-phase bulk system and ii) ice-Ih/vapor interface system where two different phases coexist. In the bulk system, the GCIceNet distinguishes nine different phases: liquid, ice-Ih, ice-Ic, ice-II, ice-III, ice-VI, ice-VII, plastic ice, and sI hydrate ice.  A benchmark study is performed to evaluate the performance of the GCIceNet relative to traditional machine learning methods. Subsequent results show that the GCIceNet outperforms all its competitors in terms of accuracy, regardless of the classification mode being supervised or unsupervised. 
 
 GCIceNet is then applied to the highly nontrivial task of identifying different phases in the ice-Ih/vapor interface system. The particular system used for the study is an ice-Ih/vapor interface, where a nanoscale quasi-liquid layer is also naturally formed at the interface. Trained in an unsupervised manner without any prior information of the phases of the water molecules, the GCIceNet performs nonlinear dimension reduction and returns relevant order parameters of the system. By applying a Gaussian mixture model to these data-discovered parameters, the Ice-Ih crystals and surface quasi-liquid layer are successfully distinguished, and hence we determine the fluctuating ice-liquid phase boundary. The relative importance analysis of the input features gives insight on how GCIceNet achieves its remarkable performance: without any prior knowledge, GCIceNet automatically finds the nonlinear mapping that assigns higher significance to the parts of the input features more suitable for describing the system. In addition to the studies of ice, GCIceNet is expected to be used in more complex multi-component systems such as amorphous and biomolecular materials.

\section*{Results and Discussion}

\subsection*{Preparation of graph structure}
\begin{figure}
\includegraphics[width=18cm]{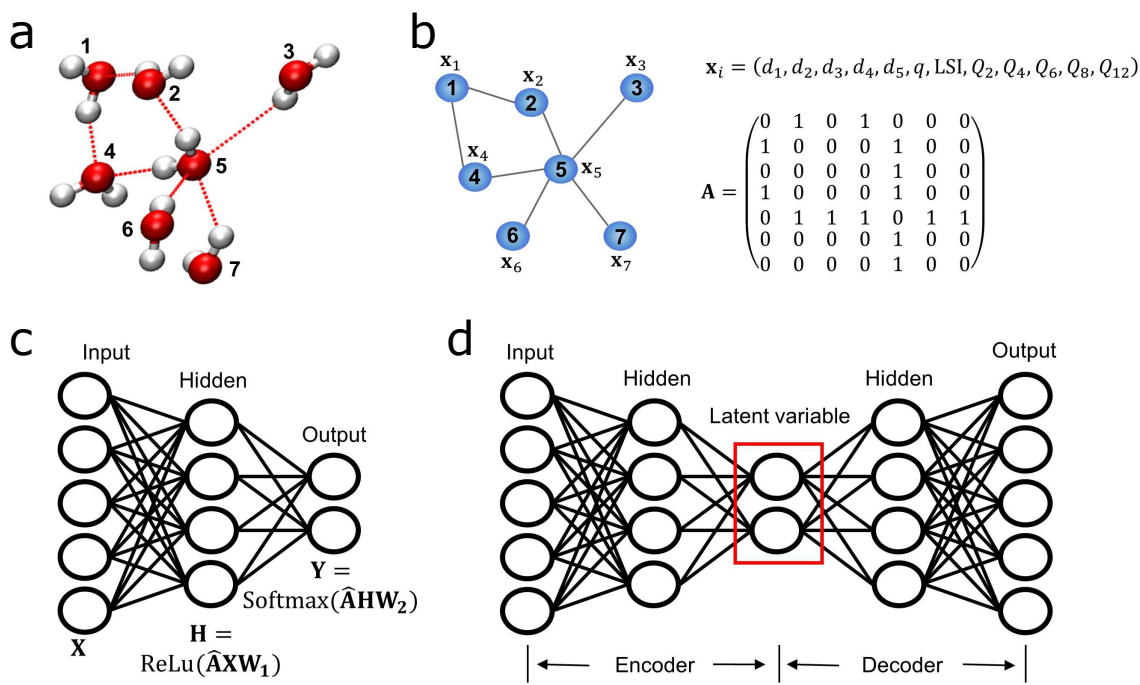}
\centering
\caption{Schematics of the graph data and graph neural networks. (a) Molecular dynamics simulation snapshot of the small water cluster. Red dotted lines denote hydrogen bonds between neighboring water molecules. (b) Graph data obtained from (a). Water molecules correspond to nodes, and hydrogen bonds correspond to edges. The feature vector $\textbf{x}_{i}$ contains order parameters of the molecule $i$. The adjacency matrix $\textbf{A}$ contains the edge information. (c) The GCIceNet structure for the supervised learning. The graph convolution is performed with the multiplication of the normalized adjacency matrix $\mathbf{\hat{A}}$. (d) The GCIceNet structure with the graph convolutional autoencoder for the unsupervised learning. It consists of symmetrical encoders and decoders, and returns latent variables from the middle layer.}
\label{fig:fig_graph} 
\end{figure}

To train the GCIceNet, we prepare the graph data representing the structural motifs of the molecules, which contains the information of the relative position and the hydrogen bonds, as shown in Fig. \ref{fig:fig_graph} (a) and (b). The graph consists of the nodes, edges, and feature vectors assigned to the nodes\cite{Trudeau1993}. The node and edge represent, respectively, the set of index assigned to each water molecule and the set of index pairs between two hydrogen-bonded water molecules. The edge is assigned only when two oxygen atoms are closer than 3.5 $\mathring{\rm{A}}$, where the first minimum of the oxygen-oxygen radial distribution function exists. When the number of water molecules is $N$, a set of edges can be represented by an adjacency matrix $\textbf{A}$ of $N \times N$ square matrix. Elements of the adjacency matrix are $A_{ij} = 1$ if the two water molecules $i$ and $j$ are connected, otherwise $A_ {ij} = 0$. The hydrogen-bond based representation of graph has been used for describing water\cite{Choi2015}, methanol\cite{Bako2013}, and osmolyte-water solutions\cite{Lee2015} and have shown that the network topology reflects the system properties. As an example, Fig. \ref{fig:fig_graph} (a) and (b) show a small water cluster obtained from the molecular dynamics simulation and its graph representation. 

A position of the water molecule $i$ is transformed into the feature vector $\textbf{x}_{i}$. It is embedded in the $i$th node. The feature vector contains the parameters conserved under translational or rotational transformation of an axis. We construct the 12-dimensional feature vector by selecting the human-made order parameters employed in the previous studies, such as the neighbor distances $ d_1 $, $ d_2 $, $ d_3 $, $ d_4 $, $ d_5 $, tetrahedral parameter $ q $, local structure index LSI, and bond orientational orders $ Q_2 $, $ Q_4 $, $ Q_6 $, $ Q_8 $, $ Q_ {12} $.  They represent the local translational order, rotational order, and crystal-like order of water molecules\cite{Tanaka2019}.  Definitions of parameters are provided in the follows.

The neighbor distance $d_{j}$ is a distance to $j$th closest neighbor molecule\cite{Cuthbertson2011}. The distance is defined as the oxygen-oxygen distance. In particular, $d_5$ denotes the radius of the first hydration shell because the water molecules form a tetrahedral structure. Therefore the $d_5$ is used to distinguish two local species in supercooled water with different local density near a liquid-liquid phase transition temperature\cite{Cuthbertson2011}.

The tetrahedral parameter $q$ is defined as\cite{Errington2001}:
\begin{equation}
q = 1- \frac{3}{8} \sum_{j = 1}^{3} \sum_{k = j + 1}^{4} \left(\mathrm {cos} \psi_{jk} + \frac{1}{3} \right)^{2},
\end{equation}
where $ \psi_{jk} $ is an angle between two vectors created by oxygen atoms of the central water molecule and two surrounding $j$th and $k$th closest neighbor molecules. $q = 1$ means that the four neighboring water molecules form a fully tetrahedral structure, and if the array is random then $ q = 0 $. In other words, $ q $ represents the local orientational order of the water molecule.

The LSI takes into account the local translation order and is defined by the following procedure\cite{Shiratani1998}. Determine the number of surrounding water molecules $n$ closer than 3.7 $\mathring{\rm{A}}$ from the central water molecule following $ r_1 <r_2 <\cdots <r_{n} <3.7 \mathring {\rm{A}} <r_{n + 1} $. After that, calculate the LSI as follows:
\begin{equation}
\mathrm{LSI} = \frac{1}{n} \sum_{j = 1}^{n} \left [\Delta (j)-\overline {\Delta} \right]^{2},
\end{equation}
where $ \Delta(j) = r_{j + 1}-r_j $ and $ \overline{\Delta} $ is the average value of $ \Delta(j) $.

The bond orientational order $Q_l$ is defined as the coarse-grained form\cite{Lechner2008} of Steinhardt parameter $q_{lm}$\cite{Steinhardt1983} as
\begin{equation}
q_{lm}\left(i\right) = \frac{1}{N} \sum_{j=1}^{N} Y_{lm} \left (\theta (\textrm{r}_{ij}), \phi (\textrm{r}_{ij}) \right),
\end{equation}
where $ \theta $, $\phi $ are polar angles, $ Y_{lm} $ is a spherical harmonic function of degree $ l $, order $ m $, and $ N $ is the number of neighboring molecules used to calculate the parameter, which we use $ N = 6 $. It is averaged as
\begin{equation}
Q_{lm}\left(i\right) = \frac{1}{N+1} \sum_{j=1}^{N}q_{lm}\left(j\right).
\end{equation}
$Q_{l}$ is obtained from the averaging order $m$ as,
\begin{equation}
Q_{l}\left( i \right) = \sqrt{\frac{4\pi}{2l+1} \sum_{m=-l}^{l}\left| Q_{lm}\left( i \right) \right|^{2}} 
\end{equation}

\subsection*{Graph Convolutional Network}
Based on the assigned feature vectors, the GCIceNet classifies the phase of the water molecules, which is represented by the label of the nodes in the graph, via supervised learning and unsupervised learning. The feature vector $\textbf{x} _{i} \in \mathbb{R}^{S}$ $(S = 12)$, which contains the human-made order parameters, is embedded in the node ${i}$. The graph is thus represented by the feature matrix $ \textbf{X} \in \mathbb{R}^{N \times S} $, where $N$ is the number of the nodes. 

When the feature matrix passes the graph convolutional layers in the GCIceNet, a convolution with the information of the neighboring nodes increase the performance of the classification\cite{Lecun2015, Kipf2016}. We carry out a spectral convolution using the normalized adjacency matrix according to Kipf's studies\cite{Kipf2016}. When the feature matrix $\textbf{X}$ passes the graph convolutional layer and returns a hidden matrix $\textbf{H}$, the convolution is defined as
 \begin{equation}
\mathbf{H} = \mathrm{ReLu} (\mathbf {\hat{A} XW}),
\end{equation}
where $ \mathbf {\hat {A}} = \mathbf {\tilde {D}}^{-\frac{1}{2}} \mathbf{\tilde{A}} \mathbf{\tilde{D }}^{\frac{1}{2}} $, $ \mathbf {\tilde{A}} = \mathbf{I} + \mathbf{A} $, and $ \tilde{D_{ii}} = \sum_{j} \tilde{A}_{ij}$. $\textbf{D}$ is a degree matrix. Using the normalized graph laplacian $\mathbf{\hat{A}}$ instead of the $\mathbf{A}$ denotes that the contribution of neighboring nodes during convolution is rescaled by their degree. 

  Fig. \ref{fig:fig_graph} (c) shows a schematic of the GCIceNet for the supervised learning. The graph convolutional layer is used in the hidden and output layer.  A rectified linear activation function is applied in the hidden layer, and a softmax activation function is applied to the output layer to construct the nonlinear map. The dimensions of the hidden and output matrix are $ \textbf{H} \in \mathbb{R}^{N \times 32} $, $ \textbf{Y} \in \mathbb{R}^{N \times 9 } $. The number of columns of the $ \textbf{Y} $ corresponds to the nine different phases we considered in the bulk system. The dataset obtained from the molecular dynamics simulation is divided into training, validation, and test sets at a ratio of 8:1:1. A dropout with a rate of 0.1 is applied to the hidden layer to prevent overfitting. The network is optimized with epochs over $10^4$.
 
 For unsupervised learning we use a graph convolutional autoencoder structure (Fig. \ref{fig:fig_graph} (d)). The graph convolutional autoencoder consists of two symmetrical graph convolutional multilayer networks, which are defined as encoder and decoder, respectively. Training of the autoencoder is related to the dimensionality reduction technique with nonlinear activation. While the encoder compresses the input data and the decoder restores it, the encoder learns to remove useless dimensions of the data with low variance and captures only important part of data. The compressed input data from the encoder is represented as the latent variables $ \mathbf{L}$. The encoder and decoder contains three hidden layers, and dimensions of the hidden and latent variable layer are $ \textbf{H}_{1}\in \mathbb{R}^{N \times 64} $, $ \textbf{H}_{2}\in \mathbb{R}^{N \times 32} $, $ \textbf{H}_{2}\in \mathbb{R}^{N \times 16} $, $ \textbf{L} \in \mathbb{R}^{N \times 2} $.  The network is optimized with at least $ 10^4 $ epochs. Notice that the autoencoder structure we use decodes the feature matrix of the input graph $\mathbf{X}$. This structure can learn more information than previously suggested autoencoders, which learn only the adjacency matrix $\mathbf{A}$\cite{Kipf2016-2}. 
 
To compare the performance of the GCIceNet, we prepare a general dense network without convolution and linear machine learning model as baseline models. The dense network uses only the feature matrix $\textbf{X}$, which is the same as the graph convolutional network with $ \mathbf {\hat{A}} = \textbf{I} $. Linear algorithms are similar to the graph convolutional networks without the nonlinear activation functions and the adjacency matrix. Support vector machine (SVM) and principal component analysis (PCA) is used for linear supervised and linear unsupervised learning, respectively. We use scikit-learn package to implement the linear algorithm, and the algorithms involving the neural network are implemented using PyTorch package\cite{Paszke2017}.

\subsection*{Classification of bulk water phases}
\begin{figure}
\includegraphics[width=18cm]{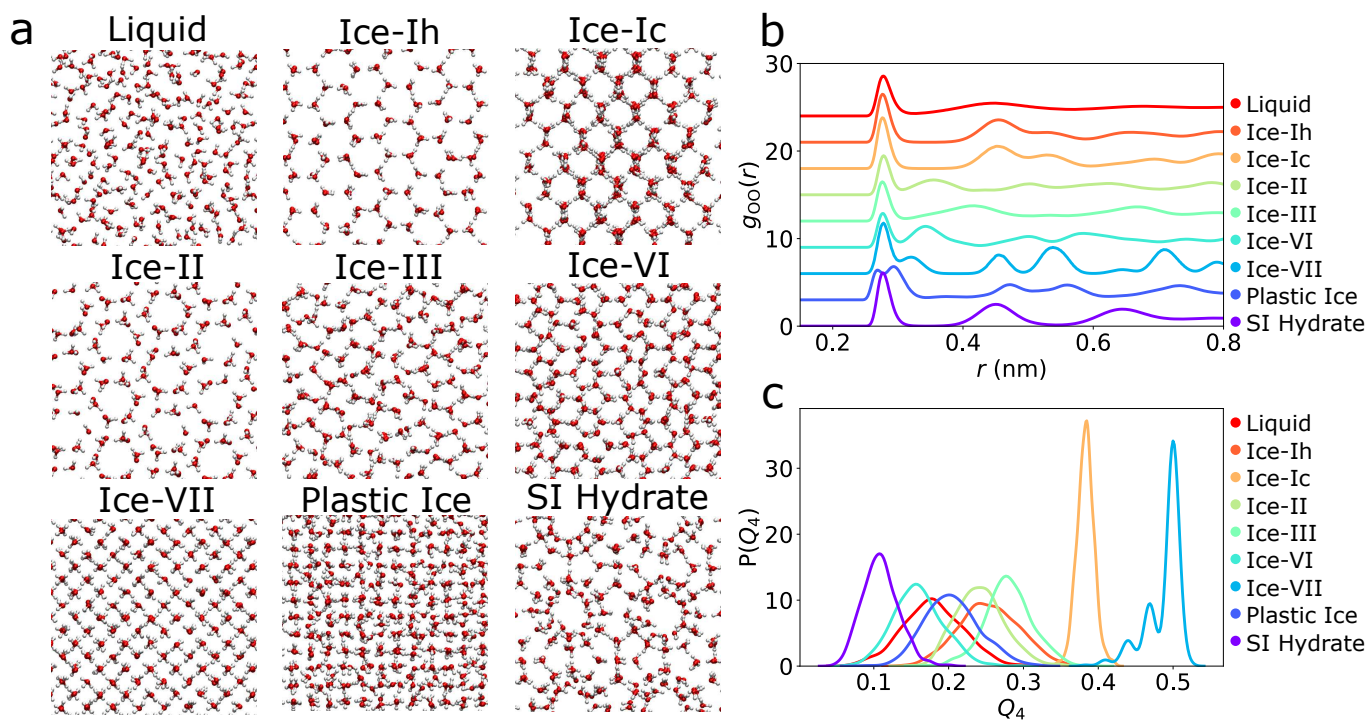}
\centering
\caption{The preparation of bulk systems for classification. (a) Molecular dynamics simulation snapshots of the liquid, ice-Ih, ice-Ic, ice-II, ice-III, ice-VI, ice-VII, plastic ice, and sI hydrate prepared for bulk system studies. Red dots are oxygen atoms, and white dots are hydrogen atoms. (b) Radial distribution functions of nine different phases between oxygen atoms. For the sake of clarity, graphs are shifted vertically. (c) Distribution of $ Q_{4} $  of the nine different phases.}
\label{fig:fig_bulk}
\end{figure}

Snapshots of nine bulk phases for the classification with GCIceNet are shown in Fig. \ref{fig:fig_bulk} (a). The nine phases consist of one liquid phase with a disordered structure, and eight ice phases with long-range order, each with different crystalline symmetries. The ice structures used in the study are from the simple hexagonal form of ice-Ih to the complex cage structure with 46 water molecules of sI hydrate. The former corresponds to the naturally occurring form of ice, and the latter occurs when compounds such as methane or carbon dioxide are in the water under high pressure. Since the temperature of the system is higher than 0 K, the snapshots of the ice phases deviate from their perfect crystalline arrangements due to thermal fluctuation-induced local vibration of the water molecules. 
 
  At first, we try to obtain a microscopic detail of molecular ordering with a radial distribution function. Fig. \ref{fig:fig_bulk} (b) shows the radial distribution functions between oxygen atoms of the water molecules of different bulk phases. For the sake of clarity, the graphs are shifted vertically. For ice crystals, the radial distribution functions exhibit long-range order. In the long range ($ r $ > 0.5 nm), each radial distribution function shows well-characterized maxima and minima. However, in the short range ($r$ < 0.5 nm), every function shows a first maximum at $ r $ = 0.28 nm and no differences are shown. It evidences that the radial distribution function, albeit exhibiting global structural information, is not suitable for determining the phases of single molecules with local information.
 
 The local structural properties of a single molecule are represented in the human-made order parameters. But the classification of phases with only the single order parameter is still incorrect. Figure \ref{fig:fig_bulk} (c) shows the bond orientational order $Q_ {4}$ distributions of the nine bulk systems. The peak positions of the liquid water, ice-Ih, and ice-Ic are all different, and minor overlaps exist between their distributions. It indicates that $Q_{4}$ can effectively distinguish these three phases. However, $Q_4$ is no longer a good measure when characterizing ice-Ih, ice-II, ice-III, ice-VII, ice-VI, and plastic ice, because there are significant overlaps between the $Q_{4}$ distributions these phases, which arise from the similarity of their symmetries. For example, ice-II, formed by compressing of ice-Ih at a temperature of 198 K at 300 MPa, still has a six-membered ring in its unit cell\cite{Bauer2008}. Not only $Q_{4}$, but also the other eleven order parameters used in the feature vector cannot effectively distinguish between the different bulk phases, as shown in Supplementary Fig. 1.
 
 \begin{table}
   \centering
    \caption{Classification accuracies of the bulk phases with three supervised learning algorithms: i) support vector machine (SVM), ii) dense network, and iii) GCIceNet.}
    \label{table:supervised_network}
 \begin{tabular}{c c c c}
          \hline
          \hline
          $$ &  SVM & Dense Network & \textbf{GCIceNet} \\
          \hline
          \hline
          Accuracy & 92.7 & 93.6 & \textbf{99.8}\\     
          \hline
 \end{tabular}
 \end{table}
 
To achieve a more accurate classification, we prepare the GCIceNet supervised networks (Fig. \ref{fig:fig_graph} (c)) and train networks with the feature vector consisting of 12 local order parameters. During the training, the neural network generates the new order parameter represented as the nonlinear mapping between the feature vector and its phase. When classifying the test dataset with the trained GCIceNet, its accuracy is 99.8 $ \% $ (Table \ref{table:supervised_network}), which is superior to the performance of the radial distribution functions and the human-made order parameters considered earlier. To compare the performance of the GCIceNet with other baseline methods, we use the dense network and support vector machine. The accuracy of the GCIceNet shows a substantial improvement compared to the 92.7 $ \% $ of support vector machine algorithm and 93.6 $ \% $ of dense neural networks without convolution. Comparing with the dense neural networks, the GCIceNet contains the graph convolutional layers and gathers information of neighboring nodes from the adjacency matrix. It indicates the importance of the graph convolutional layer for increasing the classification accuracy. Comparing with other previously studied neural network structures, the GCIceNet shows the higher accuracy than 99.6 $ \% $ of PointNet \cite{DeFever2019}, DeepIce \cite{Fulford2019}, and 98 $\%$ of Geiger-Dellago network\cite{Geiger2013} with relatively simple structure.
 
\begin{figure}
\includegraphics[width=18cm]{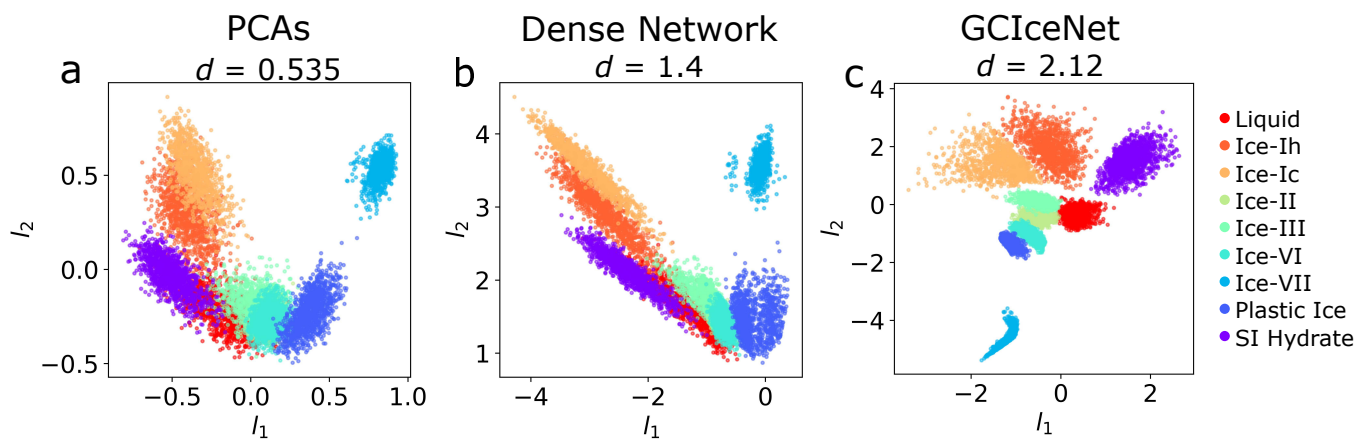}
\centering
\caption{The classification of water molecules of nine different phases with the unsupervised network. Distributions of two-dimensional latent variables are obtained with the (a) principal component analysis (PCA), (b) dense neural network, and (c) graph convolutional autoencoder of GCIceNet. Dots are colored according to labeled configurations. To quantitatively evaluate the classification performance, we show $ d $, which is defined as an average distance between the clusters of different colors.}
\label{fig:fig_bulk_unsupervised}
\end{figure}

The simple structure of GCIceNet can be easily extended to the unsupervised networks by incorporating an autoencoder architecture. Here, for unsupervised learning, we use the graph autoencoder (Fig. \ref{fig:fig_graph} (d)). Without data labels, the autoencoder trains to differentiate between input data of water phases $ \textbf{x}_{i} $ and compresses them to two-dimensional latent vectors $ \textbf{l}_{i}$, which contains the information with high variance for the unsupervised classification. Figure \ref{fig:fig_bulk_unsupervised} (c) shows the distributions of the latent vectors of bulk water systems obtained from the GCIceNet and two other baseline methods, dense network, and principal component analysis. For visual clarity, the dots are colored according to the phase label. The GCIceNet shows the best performance for the distinction of different ice phases among the three results. The linear principal component analysis hardly distinguishes between ice-II, ice-III, and ice-VI. And the dense network without the edge information suffers from overlaps between neighboring clusters except Ice-VII. The GCIceNet shows only minor overlaps between ice-II, ice-III, and ice-VI. Distances between the clusters of the different colors can be used to evaluate the performance of the unsupervised network. We introduce $ d $, which is defined as an average of the distances between the center of the nine clusters. $ d_{\textrm{GCIceNet}} $ = 2.12 and it is larger than the $ d_{\textrm{PCA}} $ = 0.535 and $ d_{\textrm{Dense}}$ = 1.4. It shows the GCIceNet shows the superior performance of the unsupervised classification as well as of the supervised classification.
 
\begin{figure}
\includegraphics[width=8.8cm]{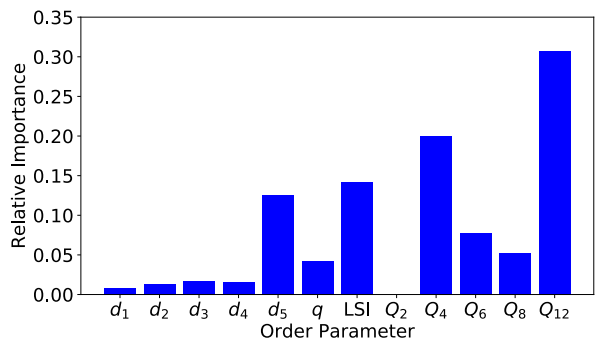}
\centering
\caption{Relative importance of the order parameters used for classification of bulk phases.}
\label{fig:fig_rip_ice}
\end{figure}

  When generating the nonlinear map, the GCIceNet assigns unbalanced weights to the elements of the 12-dimensional feature vector. The magnitude of weights indicates the importance of each element to classify the system. To estimate the element importance, we use the relative importance (RI) \cite{Boattini2019}. RI of $j$th element is defined by $\Delta \textrm{loss}_{j} $, which is a difference of a learning loss between the original dataset and new dataset by replacing the value of the $j$th element with an average of the entire dataset. Then, $\textrm{RI}_{j} $  can be defined as:
  \begin{equation}
  \textrm{RI}_{j} = \frac{\Delta\textrm{loss}_{j}} {\sum_{j} \Delta \textrm{loss}_{j}}.
  \end{equation}
  As the importance of the $j$th element increases, the $ \Delta \textrm{loss}_{j}$ increases, and the relative importance approaches 1. The relative importance of the order parameters for bulk system classification is shown in Fig. \ref{fig:fig_rip_ice} and most important parameters are $Q_{12}$, $ Q_{4} $, LSI, and $ d_{5} $. The highest weights applied in the bond orientational orders for classifying bulk phases coincides with previous studies showing that $Q_{12}$ is sensitive to the symmetry of crystal systems\cite{Errington2003, Keys2007} and $Q_{4}$ can be used to distinguish liquid and ice phases with $Q_{6}$\cite{Geiger2013}. Notice that the translational descriptors LSI and $ d_{5} $, which are rarely used in ice studies, have high relative importance. Because the bond orientational orders do not contain inter-molecular distance informations, they are compensated from the LSI and $d_{5}$ to increase the classification accuracy.

\subsection*{Clustering of Ice-Ih/vapor System}
\begin{figure}
\includegraphics[width=18cm]{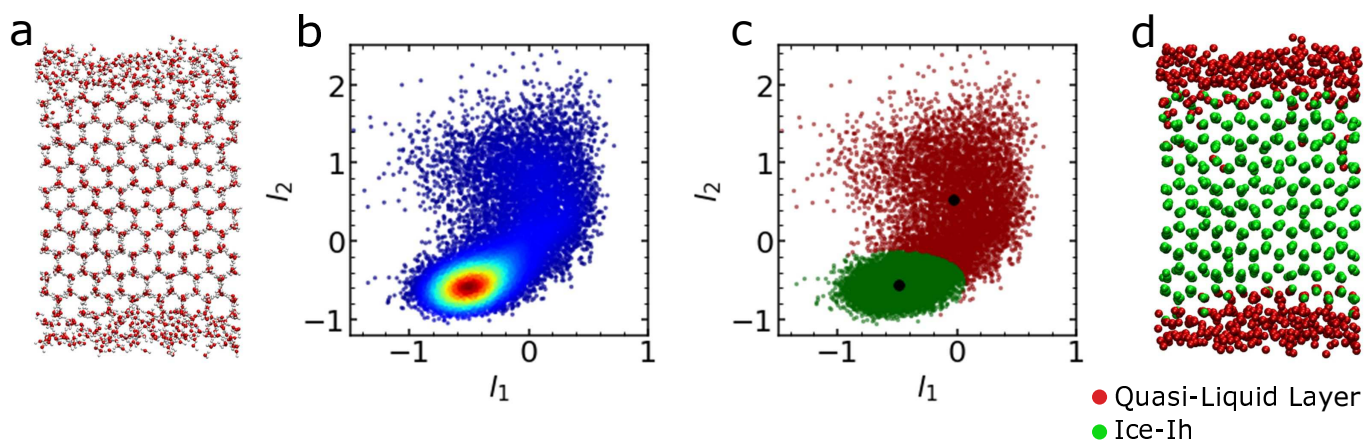}
\centering
\caption{The procedure of classification of ice-Ih and quasi-liquid layer using an unsupervised network of GCIceNet. (a) Molecular dynamics simulation snapshot of a quasi-liquid layer formed between the ice-Ih and vapor interface at $ T $ = 270 K. (b) Distribution of two-dimensional latent variables compressed with graph convolutional autoencoder of the GCIceNet. We use the kernel density estimation plot for a clear representation the of the density distribution of data. (c) Clustering of the latent variables into two groups using Gaussian mixture model. The two groups correspond to the (red dots) quasi-liquid layer and (green dots) ice-Ih respectively. (d) Re-colored snapshot of (a) based on the result of (c). Notice that the GCIceNet can recognize the quasi-liquid layer formed at the ice-Ih/vapor interface.}
\label{fig:fig_qll}
\end{figure}

 It has been experimentally\cite{Tarek2000} and theoretically\cite{Laage2009, Duboue-Dijon2015} studied that the water molecules in interfaces lose their tetrahedral ordering and show new structural and dynamical properties. As a model system we prepare an ice-Ih/vapor interface (Fig. \ref{fig:fig_qll} (a)). When the ice surface is exposed to air, a quasi-liquid layer, which is a thin film of disordered water molecules with dangling OH bonds, forms on the ice-Ih/vapor interface. An X-ray scattering study\cite{Kouchi1987} revealed the existence of the liquid layer on the ice interface, and a sum-frequency generation vibrational spectroscopy shows the disordering of water molecules in the ice interface\cite{Wei2002}. Shear viscosity measurements have shown that diffusion of the quasi-liquid layer determines the friction of the ice surface\cite{Canale2019}. Here we use the GCIceNet to separate the quasi-liquid layer molecules from the ice molecules.
 
From the positions of the water molecules of the ice/vapor system (Fig.\ref{fig:fig_qll} (a)), the graph data is prepared and trained with the unsupervised network of the GCIceNet (Fig. \ref{fig:fig_graph} (d)). The input data is compressed to the two-dimensional latent variables $ l_{1} $ and $ l_{2} $, and their distribution is shown in Fig. \ref{fig:fig_qll} (b) with kernel density estimation. The latent variables distribution can be divided into two groups: a narrow Gaussian distribution centered on the local maximum point and other data points with a broad distribution. These two groups are divided by using the Gaussian mixture model (Fig. \ref{fig:fig_qll} (c)). Notice that the centers of two clusters separated with the Gaussian mixture model similar to the density maximum estimated from the kernel density estimation. Among the two clusters, we match the cluster with narrow distribution (green dots) to the ice-Ih and the cluster with broad distribution (red dots) to the quasi-liquid layer. Figure \ref{fig:fig_qll} (d) shows the re-coloring of oxygen atoms in Fig. \ref{fig:fig_qll} with the clustering result (\ref{fig:fig_qll} (c)). It shows that the GCIceNet can extract the quasi-liquid molecules from the ice/vapor interface system.
 
\begin{figure}
\includegraphics[width=8.8cm]{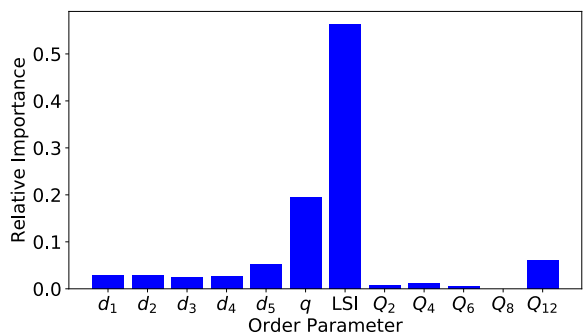}
\centering
\caption{Relative importance of the order parameters for classification of the ice-Ih and the quasi-liquid layer at $T$ = 270 K.}
\label{fig:fig_rip_qll}
\end{figure}

 Fig. \ref{fig:fig_rip_qll} shows the relative importance of the Ice-Ih/quasi-liquid layer system. Here, the translational order parameters, LSI and $q$ are important.  Relative importance of the order parameters used for the classification of the bulk (Fig. \ref{fig:fig_rip_ice}) and ice-Ih/vapor system (Fig.\ref{fig:fig_rip_qll}) are different. Because the $Q_{l}$ is developed with an assumption that the system is bulk and isotropic, it unsuccessfully describes the quasi-liquid layer with 1 nm thickness. Instead of $Q_{l}$, the tetrahedral order parameter $q$ is used to supply the information of the orientational order. Notice that the GCIceNet can learn the different properties of the bulk and ice/vapor systems only from the input data, without any prior scientific knowledge about the given system and numerous tries of the parameter combinations from the human labor.
 
\begin{figure}
\includegraphics[width=18cm]{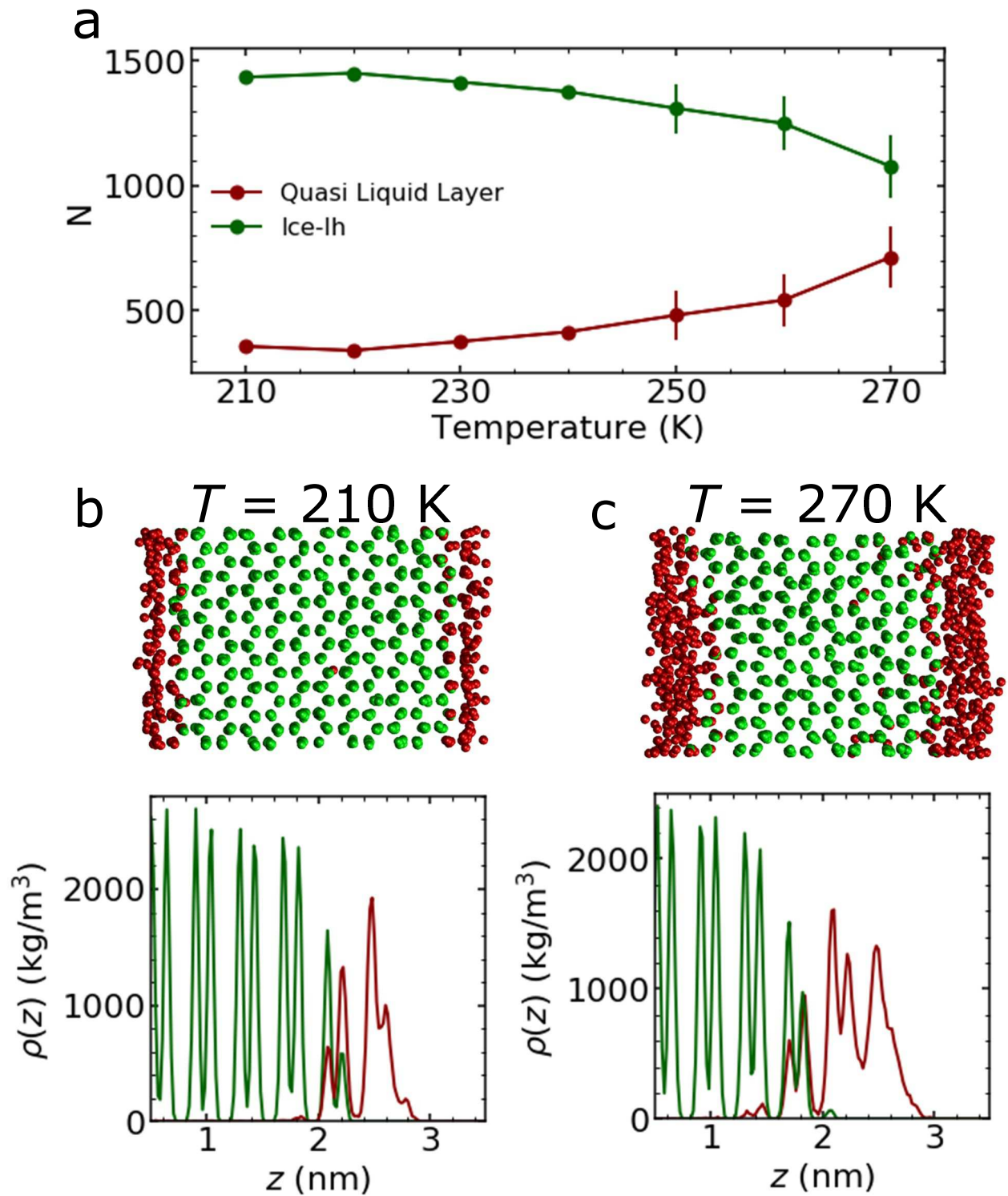}
\centering
\caption{Properties of the quasi-liquid layer with varying temperature. (a) Temperature-dependent changes in the number of water molecules of the quasi-liquid layer and ice-Ih classified by GCIceNet. As the temperature increases, the number of the quasi-liquid layer molecules increases. (Top row) Snapshots and (bottom row) density distributions of the quasi-liquid layer and ice-Ih at (b) $ T $ = 210 K, (c) $ T $ = 270 K.}
\label{fig:fig_qll_number}
\end{figure}

To show that the GCIceNet can capture the change of thermal properties of the ice/vapor system, we carry out seven additional simulations with varying temperatures from $ T $ = 210 K to $ T $ = 270 K with 10 K increments. Fig. \ref{fig:fig_qll_number} (a) shows the average number of water molecules of the quasi-liquid layer and ice-Ih classified from the GCIceNet as a function of the temperature. The number of the quasi-liquid layer molecules increases as the temperature increases, which fits well with a previous result by sum-frequency generation vibrational spectroscopy\cite{Wei2002}. The relatively large error bars at $T$ = 250 K, $T$ = 260 K, and $T$ = 270 K show an effect of thermal fluctuations on the liquid molecules at the high temperature. Figures \ref{fig:fig_qll_number} (b) and (c) show molecular dynamics snapshots and density plots of the ice-Ih and the quasi-liquid layer classified from the GCIceNet at $ T $ = 210 K and $ T $ = 270 K, respectively. The molecular dynamics snapshots and density distributions show an increase of the thickness and the disorder of the quasi-liquid layer at $ T $ = 270 K compared to $ T $ = 210 K. At $z$ = 2 nm, the density profiles of quasi-liquid layer and ice overlaps, which indicates that the interface between them is not flat. The non-flat interface shown in the instantaneous snapshot shows that the vertical position $ z $ is incomplete to define the location of the ice-Ih/quasi-liquid layer boundary. 
 
\begin{figure}
\includegraphics[width=18cm]{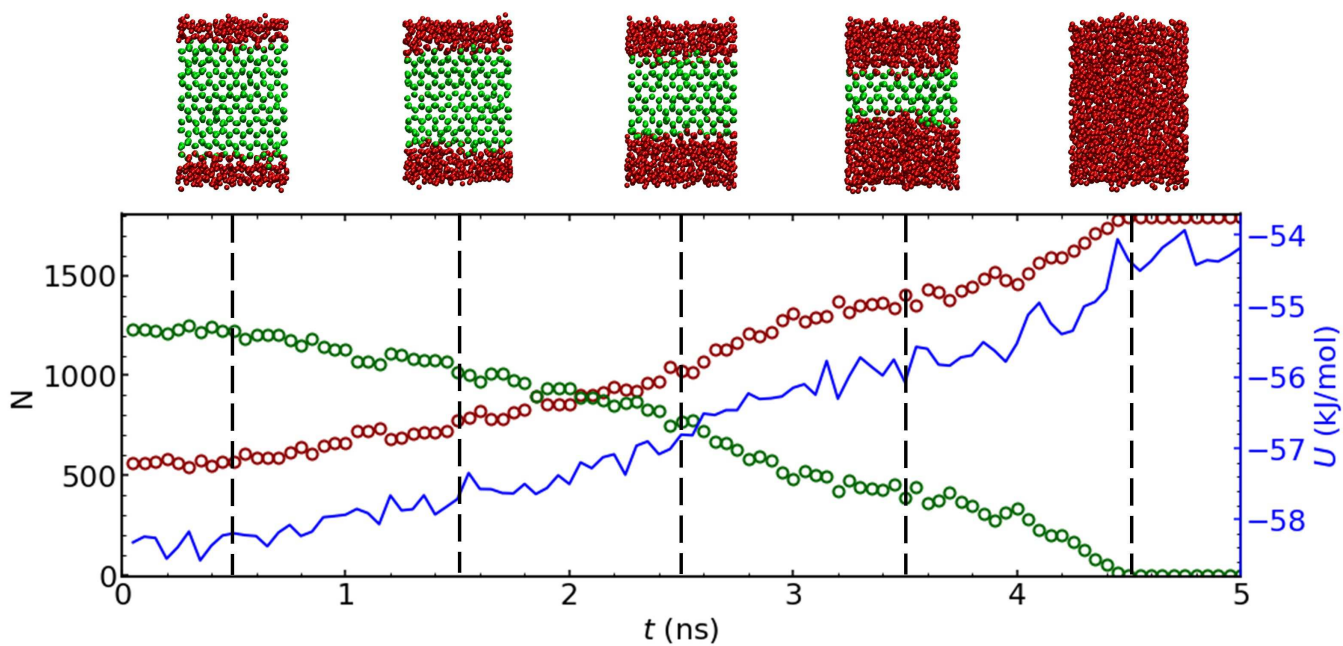}
\centering
\caption{Classification of the quasi-liquid layer and ice-Ih during the melting transition. The melting of the ice-Ih crystal, which occurs when $ T $ = 280 K, is classified into the liquid (red dots) and ice-Ih (green dots) using GCIceNet as a function of time. The potential energy of the system (blue line) is plotted for comparison. Insets show the molecular dynamics simulation snapshots classified with liquid (red dots) and ice-Ih (green dots) during the melting transition. }
\label{fig:fig_qll_melt}
\end{figure}

When the temperature of the system is higher than the melting temperature $ T_{\textrm{m}}$ = 271 K of TIP4P/Ice model\cite{Abascal2005}, the melting transition of ice-Ih initiates. Fig. \ref{fig:fig_qll_melt} shows the time series of the number of liquid and ice molecules classified by GCIceNet during the melting transition at $T$ = 280 K. The melting occurs within 5 ns, which is indicated by the zero number of ice molecules near $t = 5$ ns. To evaluate the accuracy of the GCIceNet, we compare the potential energy of the system with the number of molecules. The increasing potential energy corresponds with the increase in the number of liquid molecules, which shows that the GCIceNet can capture the path of the phase transition. Fig. \ref{fig:fig_qll_melt} shows molecular dynamics snapshots, which shows that the melting first occurs in the ice/vapor interface and progresses to the center. In other words, the melting transition of the water-vapor system is not homogeneous, but the heterogeneous phase transition where the interface acts as a nucleus.

\section*{Conclusion}

We show that the graph convolutional layers of the GCIceNet enhance the classification accuracy of the phase of water molecules. Because a regular lattice cannot contain the configuration of the water molecules, we use the graph structure with nodes, edges, and embedded feature vectors to represent the system. The graph convolutional layers in the GCIceNet use spectral properties of the graph from the feature matrix and adjacency matrix. From the molecular dynamics simulations, we prepare the bulk and ice-Ih/vapor interface system to evaluate the performance of the GCIceNet. In the bulk system, the graph convolutional layers increase the accuracy of the supervised and unsupervised classification of nine different crystalline and liquid phases. The accuracy is higher than that of other baseline algorithms, such as the dense neural network, supporting vector machine, and principal components analysis. In the ice-Ih/vapor system, the quasi-liquid layer of nanometer thickness forms between the ice and vapor. With the combination of the graph convolutional autoencoder and the Gaussian mixture clustering, the GCIceNet distinguishes the ice molecules and quasi-liquid layer molecules. In particular, the relative importance analysis shows that the GCIceNet can capture the discriminative features from the input data without any scientific prior knowledge about a given system. For example, after the training, the GCIceNet uses the bond orientational order to describe the bulk system. However, in the ice-Ih/vapor system, the GCIceNet uses the local structure index instead because the bond orientational order poorly describes the quasi-liquid layer with 1 nm thickness. With high accuracy and flexibility, the GCIceNet can be applied to other liquid crystalline systems such as colloidal liquids or liquid crystals.

\section*{Methods}
\subsection*{Molecular Dynamics Simulation}
In this letter, we use the molecular dynamics simulation to prepare the water molecules dataset. The molecular dynamics simulation numerically calculates Newton's equations of motion between molecules with the predefined force field. It returns the position and momentum of atoms as a function of the time. To model the water molecule, we use a TIP4P/Ice water model\cite{Abascal2005}. The TIP4P/Ice model is the modified version of a 4-site TIP4P model. It was developed for the phase diagram construction of the ice and amorphous water near the freezing point\cite {Abascal2005}. TIP4P/Ice model has been used to predict the homogeneous nucleation rate\cite{Espinosa2014, Haji-Akbari2015}, the binding free energy of the antifreeze protein on the ice interface\cite{Mochizuki2018}, and the shear viscosity of the ice\cite{Louden2018}. Here we carry out two kinds of simulations, a bulk system filled with a homogeneous phase and ice-Ih/vapor system where two different phases coexist.
 
 We prepare nine systems for the bulk phase simulation: liquid, ice-Ih, ice-Ic, ice-II, ice-III, ice-VI, ice-VII, plastic ice, and hydrate ice. Cage structures of the hydrate ice are filled with methane molecules in nature\cite{Kvenvolden1988}. However, we omit methane molecules and only use water molecules. 960 – 1440 water molecules fill each simulation box. In the preparation of the initial configuration ($ t = 0 $) of ice systems, we use the GenIce\cite{Matsumoto2018} package. The GenIce builds an ideal lattice structure at T = 0 K. After the energy minimization step with a steep algorithm, equilibration and production simulation are performed under the NPT ensemble, which maintains constant atomic number, pressure, and temperature. The anisotropic pressure coupling is applied to maintain pressure independently in the x-, y-, and z-direction of the rectangular simulation box. We choose the temperature and pressure of each system from the phase diagram of the TIP4P/Ice model\cite{Abascal2005} to ensure that the ice crystal does not melt during the simulation. The equilibration and production simulation run for 1 ns and 5 ns each. During the simulation, the potential energy of the system is monitored in real-time to check that no phase transition occurs.
 
In the simulation of ice-Ih/vapor systems, we prepare an ice/vapor interface. When the ice interface is exposed to the vapor, water molecules at the interface lose hydrogen bonds and form a quasi-liquid layer. The quasi-liquid layer is less than 1 nanometer thick, which determines the friction of ice surfaces measured from a stroke-probe force measurement technique\cite{Canale2019}. We create two ice/vapor interfaces by placing an ice-Ih crystal of 1792 water molecules in the middle of a $ 3.63 \times 2.96 \times 10.00 $ nm box. Two prism planes are exposed to the vapor and form liquid layers along z-direction during the simulation with the NVT ensemble. We carry out eight simulations with 10 ns equilibration and 50 ns production run, varying the temperature from 210 K to 280 K in 10 K steps to characterize the temperature dependence.
 
 The interactions between TIP4P/Ice molecules are composed of van der Waals and Coulomb interactions. We use the cutoff method with a 1.2 nm radius to calculate van der Waals interactions, and use the particle mesh Ewald algorithm\cite{Darden1993} with the 1.2 nm short-ranged cutoff to calculate Coulomb interactions. We use a leap-frog algorithm to solve Newton's equations of molecules with the 1 fs time step. Positions of atoms are saved at every 1 ps to be used in post-analysis. Temperature and pressure are controlled by a Nose-Hoover thermomstat\cite{Nose1984} and Parrinello-Rahman barostat\cite{Parrinello1981} algorithms, respectively. All simulations are performed with GROMACS 5.1.4 version package\cite{Abraham2015}.

\section*{Data availability}
The data that support the findings of this study are available from the corresponding author upon request.

\bibliography{./gcicenet.bib}

\section*{Acknowledgements}
This work was supported by the National Research Foundation of Korea (NRF) grant funded by the Korea Government (MSIP) (No. 2016R1A3B1908660).

\section*{Author contributions statement}
Q.K. and W.J. initiated the project. Q.K carried out molecular dynamics simulations. Q.K, J.K and S.K constructed GCIceNet under the guidance of W.J.. All authors participated in the discussion of the results and reviewed the manuscript.  

\section*{Competing interests}
The authors declare no competing interests.

\end{document}